\documentclass[prl,twocolumn,showpacs,preprintnumbers,superscriptaddress,amsmath,amssymb]{revtex4}

\usepackage{graphicx}% Include figure files
\usepackage{dcolumn}% Align table columns on decimal point
\usepackage{bm}% bold math
\usepackage{color}
\usepackage[colorlinks,
linkcolor=blue, urlcolor=blue, anchorcolor=blue, citecolor=blue]{hyperref}
\usepackage{multirow}

\begin{document}

\title{Parity-Enhanced Quantum Optimal Measurements}

\author{Haijun Xing }
\affiliation{Graduate School of China Academy of Engineering Physics, Beijing, 100193, China}
\affiliation{CAS Key Laboratory of Theoretical Physics, Institute of Theoretical Physics, Chinese Academy of Sciences, Beijing 100190, China}

\author{Libin Fu}
\email{lbfu@gscaep.ac.cn}
\affiliation{Graduate School of China Academy of Engineering Physics, Beijing, 100193, China}

\author{Su Yi}
\email{syi@itp.ac.cn}
\affiliation{CAS Key Laboratory of Theoretical Physics, Institute of Theoretical Physics, Chinese Academy of Sciences, Beijing 100190, China}
\affiliation{School of Physical Sciences \& CAS Center for Excellence in Topological Quantum Computation, University of Chinese Academy of Sciences, Beijing 100049, China}
\affiliation{Synergetic Innovation Center for Quantum Effects and Applications, Hunan Normal University, Changsha 410081, China}

\begin{abstract}
We find a large class of pure and mixed input states with which the phase estimation precision saturates the Cram\'{e}r-Rao bound under the compound measurements of parity and particle number. We further propose a quantum-phase-estimation  protocol for arbitrary input states, through which the precision achieved is always higher than or equal to that obtained via the original input state. We also demonstrate the implementation of the proposed scheme using a nonlinear interferometry and the realization of the nondemolition parity measurement in atomic condensates.

\end{abstract}

\maketitle

{\em Introduction}.----Parameter estimation lies at the heart of the interferometries. In typical interferometry setups, parameters are encoded into the input state of the sensor and later are inferred from the measurement results through suitable estimators~\cite{QuantMetro1,QuantMetro2,QuantMetro3}. The estimation precision is upper bounded by quantum Cram\'{e}r-Rao (CR) inequality, in terms of the quantum Fisher information (QFI)~\cite{Opt-Helstrom,Holevo,QuantMetro1,QuantMetro2,QuantMetro3}. Recently, it becomes clear that CR bound can be significantly increased by utilizing quantum resources, such as entanglement and squeezing~\cite{QuantMetro1,QuantMetro2,QuantMetro3,TFS0,TFS1,ECS1,ECS3,Caves,SSS1,SSS3,BeatSQL1,SSS2-YHX,SSS4,ECS2}. The quantum enhanced measurement precision has been experimentally demonstrated in various systems~\cite{QMetroAtom,QuantSen,QMetroOpt,AS,You2,Lee2Review,HL-Atomic} and has also been applied to the estimates of  time~\cite{clock1,clock2,clock3,clock4,clock5}, magnetic field~\cite{magn1,magn2,magn3}, and gravitational wave~\cite{GW1,GW2,GW3}, etc.

Nevertheless, saturating the upper CR bound still requires elusive optimal measurements (OMs) and suitable estimation scheme~\cite{Opt-Helstrom,Holevo,Opt-Brau,Opt-Brau2}. Generally speaking, OMs depend on the states of the system and even on the value of the parameter under estimation~\cite{Opt-Helstrom,Holevo,Opt-Brau,Opt-Brau2}. Consequently, OMs are usually achieved adaptively~\cite{Opt-Adaptive1,Opt-Zixin}. For two most commonly used measurements, parity~\cite{parity1,parity2,parity3} and number counting, in quantum metrology, there have been extensive theoretical and experimental studies on their optimality. It was shown that parity measurement is optimal for NOON state \cite{ParityHis1,ParityHis2,ParityHis3} and a few other specific states~\cite{QMetroAtom,parity3,Opt-Parity1,Opt-Parity2,Opt-Parity3}. Similar situation happens for number counting~\cite{QBYNC1,QBYNC2,QBYNC3,PS-Zhong} except for that Hofmann found a class of path-symmetric states (PSSs) which, independent of the specific phase shift, allows the CR bound to be achieved~\cite{PS-Hofmann}. However, the PSSs were only defined for pure states in conventional SU(2) interferometries~\cite{SU(2)}.

In this Letter, we identify a large class of states which, regardless of the encoded phase, achieves the CR bound under the compound measurements of parity and particle number. These states cover all PSSs and some non-PSS ones. More remarkably, they can be generalized to include the mixed states which is more relevant to experimental preparations of the input states. We further propose a complete protocol for quantum phase estimation with an arbitrary input state. The precision achieved through this protocol is always higher than or equal to that obtained via the original input state. We also demonstrate the implementation of the proposed OM scheme using a nonlinear interferometry and show the realization of the parity measurement in atomic condensates.

{\em Pure state case.}---In standard quantum metrology process~\cite{QuantMetro2}, parameter $\theta$ is encoded into the phase of a quantum state $|\psi_{\rm in}\rangle $ through a unitary transformation
\begin{flalign}
|\psi(\theta)\rangle=e^{-i\theta\hat G}|\psi_{\rm in}\rangle,
\end{flalign}
where the phase-shift generator $\hat G$ is a Hermitian operator. The eigenvalue and the corresponding eigenvector of $\hat G$ are denoted, respectively, as $g_m$ and $|m\rangle$. Furthermore, we assume that the eigenvalues satisfy $g_m=-g_{-m}$, a condition fulfilled by the widely used $\mathrm{SU}(2)$ interferometry~\cite{SU(2)} and its variants~\cite{SSS4,Lee2Review,BJJLee,Caves-nLI,Gross2010-nLI,Xing}. For convenience, we introduce the index $n\equiv|m|$. The phase-shift generator can be decomposed into
\begin{align}
{\hat G}=\sum_{n>0}g_{n}(|\uparrow\rangle_{n} {}_{n}\langle \uparrow|-|\downarrow\rangle_{n} {}_{n}\langle \downarrow|),\label{eq:genodd}
\end{align}
where $|\uparrow\rangle_n\equiv |n\rangle$ and $|\downarrow\rangle_n\equiv |-n\rangle$. The possible $|n=0\rangle$ term has been dropped in Eq.~\eqref{eq:genodd} due to $g_0=0$ based on our assumption. Moreover, as shall become clear below, this term does not contribute to the QFI and is irrelevant to the discussion about OM. Therefore, we shall always assume that $n>0$ in all summations over $n$. We note that the phase-shift operator defined by Eq.~\eqref{eq:genodd} also cover the nonlinear generators ${\hat S}_z^{3}$ \cite{NLP1} and the Ising type Hamiltonian \cite{Ising1,Ising2,Ising3}. Now, independent of the measurement operator, the precision of $\theta$'s estimator is bounded by the CR inequality
\begin{align}
\delta^2\theta\geq\frac{1}{\nu F(|\psi(\theta)\rangle, {\hat G})}
\end{align}
where $F(|\psi(\theta)\rangle, {\hat G})=4\langle \psi|{\hat G}^{2}|\psi\rangle -4\langle \psi|{\hat G}|\psi\rangle ^{2}$ is the quantum Fisher information (QFI) which measures the variance of $\hat G$ with respect to $|\psi(\theta)\rangle$~\cite{Opt-Brau} and $\nu$ is the repetitions. Clearly, achieving higher precision for the estimation of $\theta$ relies not only on the initial state which can lead to large $F(|\psi(\theta)\rangle, {\hat G})$ but also on the measurement and estimation scheme which allows the CR lower bound to be attained~\cite{Opt-Helstrom,Holevo,Opt-Brau,Opt-Brau2}. In below, by explicitly constructing orthogonal projectors, we show that for a large class of input states there exists an optimal measurement. 

To this end, we partition the Hilbert space into the direct sum of qubits with the $n$th qubit being defined by two basis states $\{|\uparrow\rangle_n,|\downarrow\rangle_n\}$. A general input state can be expanded as the superposition of qubit states, i.e.,
\begin{align}
|\psi_{\rm in}\rangle=\sum_{n}\sqrt{p_n}e^{i\varphi_n}|\alpha_n,\beta_n\rangle_n,\label{expqubit}
\end{align}
where $|\alpha_n,\beta_n\rangle_n=\cos\frac{\alpha_n}{2}e^{-i\beta_n/2}|\uparrow\rangle_n+\sin\frac{\alpha_n}{2}e^{i\beta_n/2}|\downarrow\rangle_n$ is the wave function of the $n$th qubit and $\sqrt{p_n}e^{i\varphi_n}$ is the probability amplitude with $p_n$ (subjected to the constraint $\sum_{n}p_n=1$) being the probability and $\varphi_n$ being the phase. 

The QFI of $|\psi(\theta)\rangle$ can be analytical evaluated to be
\begin{align}
F(|\psi(\theta)\rangle,{\hat G})=4\sum_{n}p_{n}g_{n}^{2} -4\bigg(\sum_{n} p_{n}g_{n}\cos\alpha_{n}\bigg)^{2},\label{qfistate}
\end{align} 
where the first and second terms originate from $\langle \hat G^2\rangle$ and $\langle \hat G\rangle^2$, respectively. For a given set of $\{p_n\}$, a sufficient condition to maximize the QFI is $\alpha_n=\pi/2$, under which each qubit lies on the equator of its own Bloch sphere. The resulting input state,
\begin{flalign}
|\psi_{E}\rangle=\sum_{n}\sqrt{p_n}e^{i\varphi_n}|\pi/2,\beta_n\rangle_n,\label{pevenstate}
\end{flalign}
is a superposition of equatorial qubits and is referred to as {\em equatorial state} (ES). As a comparison, the PSSs require that the global phase of the qubit $\varphi_n$ is independent of $n$. Therefore, $|\psi_E\rangle$ covers not only all PSSs~\cite{PS-Hofmann} but also the non-path-symmetric ones, such as the one-axis twisting spin-squeezed states \cite{SSS1,SSS4,SSS3} and the entangled coherent states \cite{ECS1,ECS2,ECS3}. More importantly, as shall be shown, the equatorial states (ESs) can also be generalized to the mixed state case.

To construct a projective measurement, we introduce a parity operator
\begin{eqnarray}
{\mathcal{P}}_{\boldsymbol 0} =\sum_{n}\left(|\downarrow\rangle_n{}_{n}\langle \uparrow|+|\uparrow\rangle_n{}_{n}\langle \downarrow|\right).\label{eq:parity}
\end{eqnarray}
It can be easily verified that ${\mathcal P}_{\boldsymbol 0}^2=1$ and the eigenvalues of ${\mathcal P}_{\boldsymbol 0}$ are $p=\pm1$. Physically, ${\mathcal P}_{\boldsymbol 0}$ inverts the spectrum of $\hat G$ as ${\mathcal P}_{\boldsymbol 0}\hat G{\mathcal P}_{\boldsymbol 0}=-\hat G$. Next, we introduce a new set of basis states for the $n$th qubit as $|{\mathbf x}^{(+)}\rangle_n\equiv|\pi/2,0\rangle_n$ and $|{\mathbf x}^{(-)}\rangle_n\equiv|\pi/2,\pi\rangle_n$,
which are of even ($p=1$) and odd ($p=-1$) parities, respectively. We then define the projection operators as
\begin{align}
\Pi_{n}^{(p)}=|{\mathbf x}^{(p)}\rangle_n{}_{n}\langle {\mathbf x}^{(p)}|,
\end{align}
satisfying $\Pi_{n}^{(p)}\Pi_{n'}^{(p')}=\delta_{nn'}\delta_{pp'}\Pi_{n}^{(p)}$ and $\sum_{p=\pm}\sum_n\Pi_{n}^{(p)}=1$. Apparently, $\{\Pi_{n}^{(p)}\}$ represents the simultaneous measurements of ${\mathcal{P}}_{\boldsymbol 0}$ and $\hat G^2$. In fact, they can be measured sequentially since $[{\mathcal P}_{\boldsymbol 0},\hat G^2]=0$. 

Now, let us perform the measurement on an ensemble of the identical states $e^{-i\theta \hat G}|\psi_E\rangle$ and denote the number of the outcomes corresponding to $\Pi_{n}^{(p)}$ after total $\nu$ repeated measurements as $\nu_n^{(p)}$. The construction of the optimal estimator can be proceeded as follows. For each set of the binary outcomes corresponding to  $\{\Pi_{n}^{(+)},\Pi_{n}^{(-)}\}$, we construct an unbiased estimator $\Theta_n$ based on the maximally likelihood estimation. The variance of $\Theta_n$ is $\delta^2\Theta_n=1/(\nu_nF_n)$ with $\nu_n=\nu_n^{(+)}+\nu_n^{(-)}$ and $F_n=4g_n^2$ \cite{ML-Fisher}. When repetition $\nu\rightarrow \infty$, we have $\delta^2\Theta_n\rightarrow 1/(\nu p_nF_n)$. Then, we take the total estimator as the linear combination of all single-qubit estimators, i.e.,
\begin{equation}
\Theta=\sum_{n}w_{n}\Theta_{n},\label{eq:estimator}
\end{equation}
where the weights $w_n$ satisfy $w_{n}\geqslant 0$ and $\sum_{n}w_{n}=1$. Apparently, $\Theta$ is still unbiased and its variance is $\delta^2\Theta=\sum_nw_n^2\delta^2\Theta_n$. It can be further shown that $\delta^2\Theta$ is minimized if $w_n=p_nF_n/F$, where $F=\sum_np_nF_n$ is the QFI of $e^{-i\theta \hat G}|\psi_{E}\rangle$. The minimal variance,
\begin{align}
(\delta^2\Theta)_{\rm min}=\frac{1}{\nu F},\label{vartheta}
\end{align}
is exactly the CR lower bound, which provers that $\{\Pi_n^{(p)}\}$ indeed represents an optimal measurement.

We comment that the optimal measurability achieved in above scheme can be attributed to the following reasons: i) For ESs, the QFI of individual qubit is maximized and the parity measurement is optimal. ii) The contributions to the total QFI from distinct qubits are decoupled [see Eq.~\eqref{qfistate}]. Hence we may perform the optimal measurement on individual qubit and construct estimator separately. iii) The weight $w_n$ in the total estimator in Eq.~\eqref{eq:estimator} is inversely proportional to $\delta^2\Theta_n$, which warrants the efficient usage of all resources. 

{\em Mixed state case.}---The pure state results can be generalized to the mixed state case straightforwardly. In order to find the desired density matrix $\rho_E$ for the input state, we recall that one of the reasons the proposed scheme works for pure states is because every qubit is an ES. Therefore, the minimum requirement for $\rho_{E}$ is that, when projected to an arbitrary qubit subspace, one should obtain an equatorial qubit, i.e., 
\begin{align}
{\Pi}_n {\rho}_{E} {\Pi}_n =|\pi/2,\beta_n\rangle_n{}_n\langle\pi/2,\beta_n|\label{eq:eqsmix}
\end{align}
for any  ${\Pi}_n\equiv {\Pi}_n^{(+)}+{\Pi}_n^{(-)}$. Correspondingly, the explicit form of the density matrix is
\begin{align}
{\rho}_{E}=&\sum_np_n|\pi/2,\beta_n\rangle_n{}_n\langle\pi/2,\beta_n|\nonumber\\
&+\sum_{m\neq n}\big(\gamma_{mn}|\pi/2,\beta_m\rangle_m{}_n\langle\pi/2,\beta_n|+\mathrm{h.c.}\big),\label{mixedst}
\end{align}
where $|\gamma_{mn}|^2\leqslant p_np_m$ due to the decoherence. This equation merely states that $\rho_E$ is supported by a {\em unique} ES of each qubit subspace.

To see that the optimal measurement can be attained with $\rho_{E}$, we evaluate the QFI of the parametrized state $\rho(\theta)=e^{-i\theta \hat G} \rho_{E} e^{i\theta \hat G}$, i.e., $F\big({\rho}(\theta), {\hat G}\big)=\mathrm{tr}({\rho}(\theta) L^2)$, where $L$ is the symmetric logarithmic derivative of ${\rho}$ that satisfies $\partial_\theta\rho(\theta)=\frac{1}{2}(L\rho+\rho L)$ and $L^\dagger=L$ \cite{Opt-Helstrom}. It can be directly verified that, in the $ G$ representation, 
\begin{align}
L=2i\sum_n g_n&\left[e^{i(2g_n\theta+\beta_n)}|\downarrow\rangle_n{}_n\langle\uparrow|+{\rm h.c.}\right]\label{SLD}
\end{align}
fulfills our purpose. Straightforward calculations give rise to $F\big(\rho_E, {G}\big)=4\sum_n p_ng_n^2$, which is again the sum of QFI of individual qubit. Now, by applying the measurement $\{ \Pi_{n}^{(p)}\}$ and constructing the same estimators $\{\Theta_n\}$ and $\Theta$ as in the pure-state case, we can also attain the minimum variance of $\Theta$ [Eq.~\eqref{vartheta}], which confirms the existence of the optimal measurement for mixed state $\rho_{E}$.

We note that $\rho_E$ may be treated as the mixed state decoheres from the pure state $|\psi_E\rangle$. The fact that  these two states have equal QFI given the same set of $\{p_n\}$ indicates that not all quantum coherence are usable for improving the precision of phase estimation. This can also be seen from the symmetric logarithmic derivative, Eq.~\eqref{SLD}, in which $\gamma_{mn}$ is completely absent. Additionally, in the construction of $\Theta$, all estimators $\Theta_n$ and weights $w_n$ are independent of $\gamma_{mn}$, which implies that the coherence between distinct qubits is irrelevant to the phase estimation.

Since the system-bath couplings that induce the decoherence are unavoidable, it is interesting to find the condition under which the optimal measurability of the input state, $|\psi_E\rangle$ or $\rho_E$, is maintained. To this end, we formally express the overall Hamiltonian (system plus bath) as
\begin{equation}
{H}=\sum_\kappa H_\kappa \otimes B_\kappa,
\end{equation}
where $H_\kappa$ and $B_\kappa$ are operators defined on the Hilbert spaces for system and bath, respectively, and $B_\kappa$ are linearly independent~\cite{Viola}. We then define a generalized state-dependent parity operator
\begin{equation}
{\mathcal{P}}_{\boldsymbol\beta}=\sum_n(e^{i\beta_n}|\downarrow\rangle_{nn}\langle\uparrow|+e^{-i\beta_n}|\uparrow\rangle_{nn}\langle\downarrow|),
\end{equation}
where $\beta_n$ are given by the state $|\psi_E\rangle$ or $\rho_E$. It can be shown that a sufficient condition for Eq.~\eqref{eq:eqsmix} being satisfied by the density matrix of the system is
\begin{align}
[H_\kappa,{\mathcal{P}}_{\boldsymbol\beta}]=0\mbox{ for any $\kappa$}.
\end{align} 
Remarkably, even if this condition is not satisfied, the optimal measurability can still be approximately preserved through dynamical decoupling~\cite{Viola,ZRGong}. In fact, by noting that ${\mathcal{P}}_{\boldsymbol\beta}$ is a unitary operator, we introduce the so-called  ${\mathcal{P}}_{\boldsymbol\beta}$ pulse which transforms a state of the system according to $\rho\rightarrow {\mathcal{P}}_{\boldsymbol\beta}\rho{\mathcal{P}}_{\boldsymbol\beta}$. Then, by applying a sequence of ${\mathcal{P}}_{\boldsymbol\beta}$ pulses with a sufficiently small inter-pulse interval, the time evolution of the system and bath is driven by the effective overall Hamiltonian $\bar{H}=\sum_\kappa\bar{H}_\kappa\otimes{B}_\kappa$, where $\bar{H}_\kappa=\frac{1}{2}\left[H_\kappa +{\mathcal{P}}_{\boldsymbol\beta}H_\kappa{\mathcal{P}}_{\boldsymbol\beta}\right]$. Clearly, the optimal measurability is maintained since $[\bar{H}_k,{\mathcal{P}}_{\boldsymbol\beta}]=0$ for any $\kappa$. We comment that the possible scenarios for applying dynamical decoupling include the input state preparation and the state storage, for which the system is very likely exposed to environment.

{\em Parity-enhanced phase-estimation scheme.}---In addition to being used for measurement and estimation, parity measurement also increases the QFI of the input state. To see this, we consider a general state $\rho$ whose QFI satisfies the inequality
$F(\rho,G)\leq 4{\rm tr}(\rho \hat G^2)-4{\rm tr}(\rho \hat G)^2$. After performing the  parity measurement ${\mathcal P}_{\boldsymbol 0}$ on $\rho$, the state collapses into the ESs
\begin{align}
\rho^{(\pm)}=\Pi^{(\pm)}\rho\Pi^{(\pm)}/q^{(\pm)},
\end{align}
where $\Pi^{(\pm)}=\sum_n\Pi_n^{(\pm)}$ are projections to the even- and odd-parity subspaces, respectively, and $q^{(\pm)}={\rm tr}(\rho\Pi^{(\pm)})$ are the probabilities to obtain the outcomes $\pm1$. The average QFI of the resulting states is
\begin{align}
\bar F=\sum_{p=\pm}q^{(p)}F(\rho^{(p)})=4{\rm tr}(\rho \hat G^2)\geq F(\rho,G),\label{meanf}
\end{align}
which indicates that the measuring ${\mathcal P}_{\boldsymbol 0}$ indeed improves the quality of the input state.

\begin{figure}[ptb]
\centering
\includegraphics[width=0.95\columnwidth, trim=0 0 400 387,clip]{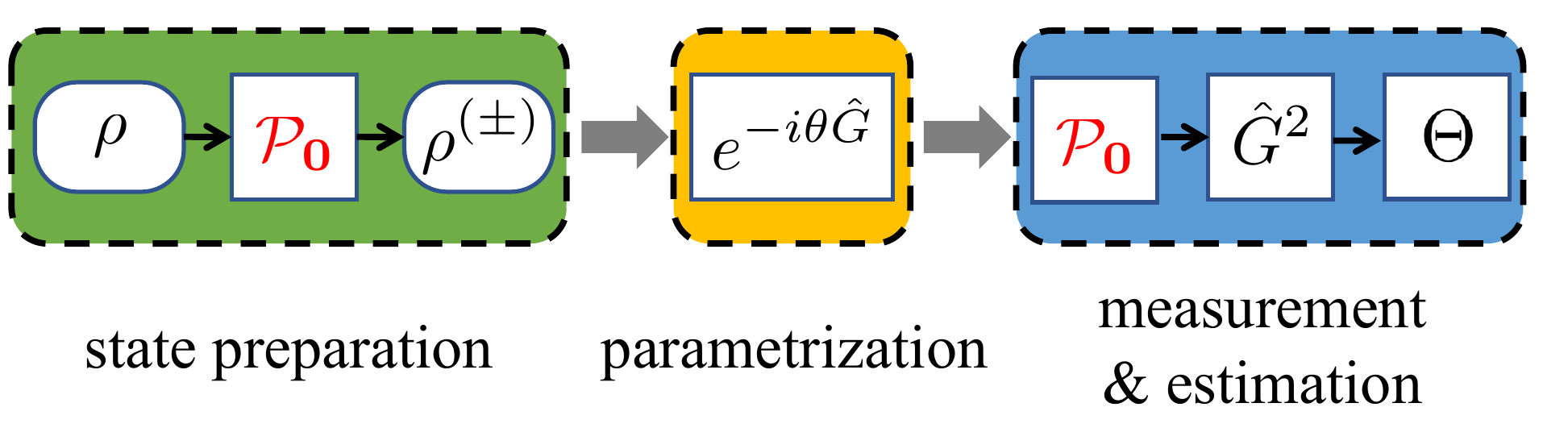}
\caption{(color online). Schematics for the protocol of optimal phase estimation. i) An ES can be prepared directly or via a parity measurement ${\mathcal P}_{\boldsymbol 0}$. ii) A phase $\theta$ is encoded by the generator $\hat{G}$. iii) CR bound of the phase estimation precision is attained with the measurements ${\mathcal P}_{\boldsymbol 0}$ and $\hat{G}^2$ and with the estimator $\Theta$.}
\label{figope}
\end{figure}

In Fig.~\ref{figope}, we schematically summarize the protocol for optimal phase estimation. Interestingly, for state preparation, if the input state is an eigenstate of $\hat G$, say $|\uparrow\rangle_n$, a parity measurement would yield an equatorial state, $|{\mathbf x}^{(\pm)}\rangle_n$, in the $n$th qubit subspace. Correspondingly, the QFI of the input state is increased from zero to $4g_n^2$. In particular, the QFI is maximized if $g_n^2$ is the largest eigenvalues of $\hat G^2$. Therefore, the efficiency of parity measurement for input state preparation can be extremely high. It is worthwhile to mention that, other than being used for state preparation and measurement, parity is also useful for state storage as discussed previously. 

{\em Nonlinear interferometry}.---To demonstrate the applications of the proposed scheme, we consider an nonlinear interferometry modeled by the Hamiltonian
\begin{equation}
H_{\rm ni}(t)=-\chi \hat S_z^2 -B_x(t) \hat S_x + B_z \hat S_z,\label{BBJ-H}
\end{equation}
where, for a two-mode system, say modes a and b, the angular momentum operators are defined as $\hat S_x=(\hat a^\dag \hat b+\hat b^\dag \hat a)/2$, $\hat S_y=(\hat a^\dag \hat b-\hat b^\dag \hat a)/(2i)$, and $\hat S_z=(\hat a^\dag \hat a-\hat b^\dag \hat b)/2$ with $\hat a$ and $\hat b$ being the annihilation operators for modes a and b, respectively, and $\hat S_z$ is the phase-shift generator. Furthermore, $\chi$ $(>0)$ is the nonlinear coupling strength, $B_x$ and $B_z$ are transverse and longitudinal fields, respectively. We point out that Hamiltonian Eq.~\eqref{BBJ-H} can be realized by either external or internal states of $N$ Bose condensed atoms \cite{Xing, BJJLee, Lee2Review, Caves-nLI, Gross2010-nLI}.

To start, let us first briefly recall the eigenspectrum of $H_{\rm ni}$ in the absence of the longitudinal field $B_z$~\cite{BJJLee,Xing}. For $B_x\gg N\chi$, the eigenstates of $H_{\rm ni}$ are those of $\hat S_x$, i.e., $|m_x=m\rangle$ with $m=-N/2,-N/2+1,\ldots,N/2$. In particular, the parity of $|m_x\rangle$ is $(-1)^{N/2-m_x}$. While at the $B_x=0$, the eigenstates of $H_{\rm ni}$ are those of $\hat S_z^2$ which are doubly degenerate. By varying $B_x$, these two sets of spectra are adiabatically connected according to
\begin{align}
|{\mathbf x}^{(\pm)}\rangle_n\leftrightarrow |m_x=2n-(N+1\mp1)/2\rangle.\label{mapszsx}
\end{align}
The nonlinear interferometry is generally operated as follows. Initially, the system is prepared in state $|\Psi_0\rangle=|m_x=N/2\rangle$ under a large $B_x$. The transverse field is then swept to zero which gives rise to the input state for the interferometry $|\Psi_{\rm in}\rangle=\sum_nc_n|{\mathbf x}^{(+)}\rangle_n$, where $c_n$ depend on the sweeping rate $dB_x/dt$. Clearly, $|\Psi_{\rm in}\rangle$ is an ES with all $\beta_n=0$. We remark that $|\Psi_{\rm in}\rangle$ has the same even parity as that of $|\Psi_0\rangle$ since the Hamiltonian for input preparation converses ${\mathcal P}_{\boldsymbol 0}$. In fact, as discussed previously, even in the presence of stray fields, dynamical decoupling can be used to recover the parity conservation with ${\mathcal P}_{{\boldsymbol\beta}={\boldsymbol 0}}=(-1)^{N/2}e^{-i\pi\hat S_x}$ pulse.

To proceed further, we turn on the longitudinal field for a time interval $\Delta t$, which encodes the phase $\theta=B_z\Delta t$ into the wave function through $|\Psi(\theta)\rangle=e^{-i\Delta tH_{\rm ni}}|\Psi_{\rm in}\rangle$. We remark that since the nonlinear term $S_z^2$ only contributes a global phase, $e^{-i\Delta t\chi \hat S_z^2}$, to each qubit, $|\Psi(\theta)\rangle$ is still an ES. It, however, is not a PSS due to this phase. Finally, we adiabatically increase $B_x$ to a value much larger than $N\chi$, which maps $|{\mathbf x}^{(\pm)}\rangle_n$ back to the eigenstate of $\hat S_x$ based on Eq.~\eqref{mapszsx}. The measurement $\{\Pi^{(\pm)}_n\}$ can then be realized by measuring $\hat S_x$ with the resulting state.

We point out that measurement $\{\Pi^{(\pm)}_n\}$ relies on two conditions: i) the adiabatic energy spectrum of $H_{\rm ni}$ is nondegenerate, which is generally true unless there exist accidental degeneracies. ii) $\hat S_x$ is directly measurable by, e.g. Stern-Gerlach apparatus. Otherwise, more sophisticated approaches have to be applied~\cite{BJJLee}.

{\em Realization of nondemolition parity measurement.}---If $\{\Pi^{(\pm)}_n\}$ cannot be implemented as a single measurement, one may measure ${\mathcal P}_{\boldsymbol 0}$ and $\hat G^2$ sequentially, which requires that the measurement of ${\mathcal P}_{\boldsymbol 0}$ is nondemolition as those experimentally realized in various systems~\cite{QND_P, QND-PO,QND_Ptheo}. Here, as an example, we demonstrate its realization in a two-mode atomic system, for which the parity operator becomes 
\begin{align}
{\mathcal P}_{\boldsymbol 0}=(-1)^{S-\hat S_x}=e^{-i\pi\hat S_y/2}(-1)^{\hat b^\dag \hat b}e^{i\pi\hat S_y/2}.\label{p0to1}
\end{align}
As can be seen, other than the $\pi/2$ rotations around the $y$ axis, the measurement of ${\mathcal P}_{\boldsymbol 0}$ is reduced to that of $(-1)^{\hat b^\dagger\hat b}$ which, in analog to the parity measurement of the photon number in cavity~\cite{QND_P}, can be realized by introducing an ancilla qubit coupling to mode b of the system. Specifically, we assume the qubit-system coupling Hamiltonian takes the form
\begin{align}
H_{\rm qs}/\hbar= \omega_{q}|e\rangle\langle e|+\chi_{\rm qs}\hat b^{\dagger}\hat b|e\rangle\langle e|\label{qusys},
\end{align}
where $\hbar\omega_{q}$ is the energy difference between the ground state, $|g\rangle$, and the excited state, $|e\rangle$, of the qubit, and $\chi_{\rm qs}$ is the qubit-system coupling strength. In SM~\cite{suppl}, we show how to engineer Hamiltonian Eq.~\eqref{qusys} with the internal states of atoms. In the rotating frame of the qubit, the excited state of the qubit acquires a phase $\Phi=\chi_{\rm qs}\hat b^\dagger\hat bt$ proportional to the atom number in mode $\hat{b}$. Then by carefully choosing the evolution time $t$ such that $\chi_{\rm qs}t=\pi$, we realize the operation
$U_\pi=(-1)^{\hat b^\dagger \hat b}\otimes |e\rangle\langle e|+\hat I_s\otimes|g\rangle\langle g|$, where $\hat I_s$ is the identity operator of the system. Then by inserting $U_\pi$ between $\pi/2$ and $-\pi/2$ rotations around the $y$ axis for both qubit and system, we realize a controlled-$X$ gate
\begin{align}
C_X&= e^{-i\pi\hat S_y/2}R_y^\dag\left(\frac{\pi}{2}\right)U_\pi R_y\left(\frac{\pi}{2}\right)e^{i\pi\hat S_y/2}\nonumber\\
&=\Pi^{(+)}\otimes \hat I_q+\Pi^{(-)}\otimes \hat\sigma_x,
\end{align}
where $R_y(\pi/2)$ is the $\pi/2$ rotation of the qubit around the $y$ axis, $\hat I_q$ is the identity operator of the qubit, and $\hat{\sigma}_x=\big(|e\rangle\langle g|+|g\rangle\langle e|\big)$ flips the qubit. To perform the measurement, we may prepare the qubit in, e.g., $|g\rangle$ state, $C_X$ couples the even (odd) parity state of the system to $|g\rangle$ ($|e\rangle$). A subsequent projective measurement on the qubit will leave the system in a parity-definite state, which completes the measurement ${\mathcal P}_{\boldsymbol 0}$.

{\em Conclusions.}---We have proposed an OM scheme for the pure and mixed ESs which cover a wide range of the input states in various interferometries. Our scheme base on the combined measurement of parity and particle number, which allows us to unveil more information of the states compared to the single measurement of either one. We have also proposed a protocol for phase estimation by including the state preparation using parity measurement, which the precision achieved through our protocol is always higher than or equal to that obtained via the original input state. We also demonstrate the implementation of the proposed OM scheme using a nonlinear interferometry and show the realization of the parity measurement in atomic condensates.

HJX thanks Dr. Yingdan Wang and Dr. Stefano Chesi for the helpful discussion.  This work was supported by the NSFC (Grants No. 11434011, No. 11674334, No. 11747601, No. 11725417, No.11575027), National Key Research and Development Program of China (Grants 2017YFA0304501), NSAF (Grant No. U1730449), Science Challenge Project (Grant No. TZ2018005), and the Strategic Priority Research Program of Chinese Academy of Sciences (Grant No. XDB28000000).

%%%%%%%%%%%%%%%%%%%%%%%%%%%%%%%%%%%%%%%%%%%%%%%%%%%%%%%%%%%%%%%%%%

\onecolumngrid  %  change to single-column format
\clearpage
\renewcommand{\thesection}{S-\arabic{section}}
\setcounter{section}{0}  %  this will re-count section from 1
\renewcommand{\theequation}{S\arabic{equation}}
\setcounter{equation}{0}  %  this will re-count eq from 1
\renewcommand{\thefigure}{S\arabic{figure}}
\setcounter{figure}{0}  %  this will re-count eq from 1

\section{Supplemental Materials}

\maketitle

\subsection{I. Engineering of Hamiltonian Eq. (22)}
Here we demonstrate how to engineer the Hamiltonian Eq. (22) in the main text with an impurity qubit immersed in a two-component condensate. The coupled qubit-system Hamiltonian consists of three parts: $H=H_S+H_Q+H_I$, where $H_S$, $H_Q$, and $H_I$ describe the condensate, qubit, and qubit-system coupling, respectively. Specifically, for the condensate part, we have
\begin{align}
H_S&=\sum_{i=a,b}\int d{\mathbf r}\psi_i^\dag({\mathbf r})\left[\frac{{\mathbf p}^2}{2m_S}+{\mathcal E}_i+V_S({\mathbf r})+\frac{1}{2}\frac{4\pi\hbar^2 a_{ii}}{m_S}\psi_i^\dag({\mathbf r})\psi_i({\mathbf r})\right]\psi_i({\mathbf r})+\frac{4\pi\hbar^2 a_{ab}}{m_S}\int d{\mathbf r}\psi_a^\dagger({\mathbf r})\psi_b^\dag({\mathbf r})\psi_b({\mathbf r})\psi_a({\mathbf r}),
\end{align}
where $\psi_i({\mathbf r})$ is the field operator for the atoms in $i$th mode, ${\mathcal E}_i$ is the energy of the $i$th mode, $m_S$ is the mass of the atom, $V_S({\mathbf r})$ the external potential for condensate atoms, $a_{ii}$ the intra-species scattering lengths, and $a_{ab}$ the inter-species scattering length. For simplicity, we assume that ${\mathcal E}_a={\mathcal E}_b={\mathcal E}$ and $a_{ij}=a_S$ for any $i$ and $j$. The field operators are then simplified to $\hat \psi_a({\mathbf r})=\psi({\mathbf r})\hat a$ and $\hat \psi_b({\mathbf r})=\psi({\mathbf r})\hat b$ with $\psi({\mathbf r})$ being the mode function. The condensate Hamiltonian now reduces to
\begin{align}
H_S=({\mathcal E}'-g)N+gN^2
\end{align}
where $N=\hat a^\dagger\hat a+\hat b^\dagger\hat b$ is the total particle number operator, ${\mathcal E}'=\int d{\mathbf r}\psi^*({\mathbf r})\left[{\mathbf p}^2/(2m_S)+{\mathcal E}+V_S({\mathbf r})\right]\psi({\mathbf r})$, and $g=(4\pi\hbar^2 a_S/m_S)\int d{\mathbf r}|\psi({\mathbf r})|^4$.

Next, we turn to consider the qubit Hamiltonian which is simply
\begin{align}
H_{Q}&=\sum_{\sigma=e,g}\int d{\mathbf r}\phi_\sigma^\dag({\mathbf r})\left[\frac{{\mathbf p}^2}{2m_Q}+\varepsilon_\sigma+V_Q({\mathbf r})\right]\phi_\sigma({\mathbf r}),
\end{align}
where $\phi_\sigma({\mathbf r})$ is the field operators for the excited ($e$) and ground ($g$) states, $\varepsilon_\sigma$ are the corresponding energies, $m_Q$ is the mass of the impurity atom, and $V_Q({\mathbf r})$ is the confining potential. Generally, the trapping potential for the impurity atom is very tight such that the center of mass motion of the qubit is frozen to the ground state of $V_Q$, say $\phi({\mathbf r})$. The qubit Hamiltonian then reduces to 
\begin{align}
H_{Q}=\varepsilon_e'|e\rangle\langle e|+\varepsilon_g'|g\rangle\langle g|,
\end{align}
where $\varepsilon_\sigma'=\int d{\mathbf r}\phi^*({\mathbf r})\left[{\mathbf p}^2/(2m_Q)+\varepsilon_\sigma+V_Q({\mathbf r})\right]\phi({\mathbf r})$.

Finally, for the qubit-system coupling, we assume that only the exited state of the qubit interacts with the condensate atom in mode b with a scattering length is $a_{eb}$. Therefore, the interaction Hamiltonian is
\begin{align}
H_{I}&=\frac{2\pi\hbar^2 a_{eb}}{\bar m}\int d{\mathbf r} \psi_b^\dag({\mathbf r})\phi_e^\dag({\mathbf r})\phi_e({\mathbf r}) \psi_b({\mathbf r})=\chi_{\rm qs} |e\rangle\langle e|\hat b^\dag \hat b
\end{align}
where $\bar m=m_Sm_Q/(m_S+m_Q)$ is the reduced mass and $\chi_{\rm qs}=(2\pi\hbar^2 a_{eb}/\bar{m})\int d{\mathbf r}|\psi({\mathbf r})|^2|\phi({\mathbf r})|^2$.

Now put everything back together, we have
\begin{align}
H=({\mathcal E}'-g)N+gN^2+\varepsilon_e'|e\rangle\langle e|+\varepsilon_g'|g\rangle\langle g|+\chi_{\rm qs} |e\rangle\langle e|\hat b^\dag \hat b. 
\end{align}
After dropping the constant $N$ and $N^2$ terms and setting $\varepsilon_g'$ as the zero energy 
\begin{align}
H_{\rm qs}=\hbar\omega_q|e\rangle\langle e|+\chi_{\rm qs} |e\rangle\langle e|\hat b^\dag \hat b.
\end{align}

\end{document}